# Pure bound field corrections to the atomic energy levels and the proton size puzzle

Alexander L Kholmetskii[1], Oleg V Missevitch[2] and Tolga Yarman[3,4]

[1]Department of Physics, Belarusian State University, 4 Nezavisimosti Avenue, 220030 Minsk, Belarus
[2]Institute for Nuclear Problems, Belarusian State University, 11 Bobruiskaya Str., 220030 Minsk, Belarus
[3]Department of Engineering, Okan University, Akfirat, Istanbul, Turkey
[4]Savronik, Eskisehir, Turkey

E-mail: khol123@yahoo.com

**Abstract**

Reinforcement of the puzzle about the proton charge radius $r_E$, stimulated by the recent experiment with muonic hydrogen (A. Antognini, et al. Science **339** (2013) 417) induced a new round of discussions on the subject, and now some physicists are ready to adopt the exotic properties of muon, lying beyond the Standard Model, in order to explain the difference between the results of muonic hydrogen experiments ($r_E$=0.84087(39) fm) and CODATA-2010 value $r_E$=0.8775(51) fm based on electron-proton scattering and H spectroscopy. In the present contribution we suggest a way to achieve a progress in the entire problem via paying attention on some logical inconsistency of fundamental equations of atomic physics, constructed by analogy with corresponding classical equations without, however, taking into account a purely bound nature of electromagnetic fields generated by the electrically bound particles in the stationary energy states. We suggest eliminating this inconsistency via introducing some appropriate correcting factors into these equations, which explicitly involve the requirement of total momentum conservation in the system "bound particles and their fields" in the absence of electromagnetic radiation. We further show that this approach allows us not only to eliminate long-standing discrepancies between theory and experiment in precise physics of simple atoms, but also yields the same estimation (though with different uncertainties) for the proton size in the classic 2S-2P Lamb shift in hydrogen, 1S Lamb shift in hydrogen, and 2S-2P Lamb shift in muonic hydrogen, with the mean value $r_E$=0.841 fm. Finally, we suggest the crucial experiment for verification of the validity of pure bound field corrections: the measurement of lifetime of bound moun in various meso-atoms, especially at large $Z$, where the standard calculations and our predictions essentially deviate from each other, and some of the available experimental results (Yovanovitch, Phys. Rev. **117** (1960) 1580) strongly support our approach.

## 1. Introduction

In the recent papers by Antognini et al. [1] the authors reported the new result of measurement of the proton charge radius $r_E$ via the laser spectroscopy in muonic hydrogen ($r_E$=0.84087(39) fm), which confirms their previous result [2] with the enhanced precision. The crucial problem, which now attracts a great attention of the scientific community, is the drastic deviation between the reported result and the CODATA-2010 value $r_E$=0.8775(51) fm [3] based on electron-proton scattering and H spectroscopy.

Currently the accent is mainly made on the deviation in estimation of $r_E$ in the muonic hydrogen and in the electron-proton scattering experiments, whereas the results of measurement of the proton size in the classic Lamb shift (which give the averaged value of $r_E$=0.883(6) fm [4], being even larger than in the scattering experiments) is mentioned less frequently. However, it seems that the different estimations of $r_E$ via the Lamb shift in hydrogen and in muonic hydrogen represent the most puzzling result, because both kinds of measurements are carried out with the same (at least basically) experimental technique, the laser spectroscopy. Hence we are sure that the elimination of this discrepancy can give a key to the solution of the entire problem of the proton size puzzle. Hereinafter we do not assume that muon-proton interaction differs from electron-proton interaction, supposing this hypothesis exotic.

In this respect we refer to our recent papers [4, 5], where we developed a novel approach in the atomic physics, conditionally named as Pure Bound Field Theory (PBFT), which gives the equal values of the proton charge radius calculated via the classic Lamb shift in hydrogen ($r_E$=0.841(6) fm), 1S Lamb shift in hydrogen ($r_E$=0.844(22) fm) and muonic hydrogen ($r_E$=0.84087(39) fm). Here the indicated uncertainty mainly stems from the corresponding measurement uncertainty.

What is more, we will show below that this approach allows eliminating the available long-standing discrepancies between theory and experiment in precise physics of simple atoms, which, along with the re-estimation of the proton size, makes hardly to believe that this success is occasional.

The principal motivating factor to the development of our approach is the observation that basic equations of atomic physics, being constructed by analogy with the appropriate classical equations, ignore, however, the principal difference between electromagnetic fields of classically bound charges (which generate both bound and radiative field components) and quantum bound charges (whose fields in the stationary states contain the bound component alone).

The ignorance of such difference can be clearly seen in the Breit equation and Bethe-Salpeter equation [6] for the quantum two-body problem, which essentially use the classical analog of the law of conservation of total momentum in the system "particles and fields", expressed as

$$\boldsymbol{p}_m = -\boldsymbol{p}_M, \qquad (1)$$

where $\boldsymbol{p}_m$, $\boldsymbol{p}_M$ are the generalized momenta of particles $m$ and $M$, correspondingly.

However, due to the difference in the structure of electromagnetic fields for bound classical and quantum particles, eq. (1) cannot, in general, be straightforwardly extended from the classical to quantum domain due to the known fact that electromagnetic radiation which possesses a momentum is absent in the quantum case. Thus, as minimum, in the quantum equations the generalized momenta $\boldsymbol{p}_m$ and $\boldsymbol{p}_M$ should be re-defined in the way, which maintains the total momentum conservation law in the absence of momentum component, associated with an electromagnetic radiation.

In order to solve this problem, in refs. [5, 7] we considered a semi-classical limit of the two-body problem with the prohibited radiation of bound charges. This way we have shown that the total momentum conservation law holds, when the generalized momenta in eq. (1) are redefined as:

$$\boldsymbol{p}_m = \gamma_m m \boldsymbol{v}_m + \gamma_m \frac{M}{M+m}\frac{U}{c^2}\boldsymbol{v}_m = \gamma_m b_m m \boldsymbol{v}_m, \qquad (2a)$$

$$\boldsymbol{p}_M = \gamma_M M \boldsymbol{v}_M + \gamma_M \frac{m}{M+m}\frac{U}{c^2}\boldsymbol{v}_M = \gamma_M b_M M \boldsymbol{v}_M, \qquad (2b)$$

where $\boldsymbol{v}_m$, $\boldsymbol{v}_M$ are the velocities of particles $m$ and $M$, correspondingly, $\gamma_m$, $\gamma_M$ are their Lorentz factors

$$\gamma_m = \left(1-\frac{v_m^2}{c^2}\right)^{-1/2} = \left[1+\frac{U}{mc^2}\frac{M^2}{(m+M)^2}\right]^{-1/2}, \quad \gamma_M = \left(1-\frac{v_M^2}{c^2}\right)^{-1/2} = \left[1+\frac{U}{Mc^2}\frac{m^2}{(m+M)^2}\right]^{-1/2}, \qquad (3\text{a-b})$$

and $U$ is the interactional electromagnetic energy of bound particles. Here we also introduced the definitions

$$b_m = \left(1+\frac{U}{mc^2}\frac{M}{M+m}\right), \quad b_M = \left(1+\frac{U}{mc^2}\frac{m}{M+m}\right), \qquad (3\text{c-d})$$

which can be named as the binding factors of particles $m$ and $M$, correspondingly. Thus, formally eqs. (2a-b) can be obtained via the replacement of the rest masses of particles $m$ and $M$ by their effective values

$$m \to b_m m, \ M \to b_M M . \qquad (4a\text{-}b)$$

In addition, we have shown [5] that for the bound (velocity-dependend) electromagnetic fields of both particles, given by the Heaviside solution of Maxwell equations (see, e.g. [8]), the total momentum conservation law requires the replacement of electromagnetic interactional energy by its effective value

$$U \to \gamma_m \gamma_M U \qquad (5)$$

for the circular motion of two particles around their common center of mass.

The next problem is to determine the quantum counterparts of factors $\gamma_m$, $\gamma_M$, $b_m$, $b_M$, which, as seen from eqs. (3), should be the constant values in the stationary energy states. Further on we suppose that similar factors should be introduced for any discrete energy level $n$, so that $\gamma_{mn}$, $\gamma_{Mn}$, $b_{mn}$, $b_{Mn}$ occur the functions of principal quantum number $n$ only[1]. Hence, the introduction of these factors into the basic equations of atomic physics do not affect the Lorentz invariance of these equations, as well as other known symmetries of electromagnetic interaction.

In order to determine the factors $\gamma_{mn}$, $\gamma_{Mn}$, $b_{mn}$, $b_{Mn}$ explicitly, we introduced the replacements similar to (4), (5) into the Breit equation without external field [6] for quantum two-body problem, i.e.

$$U \to \gamma_{mn}\gamma_{Mn}U, \ m \to b_{mn}m, \ M \to b_{Mn}M, \qquad (6a\text{-}c)$$

and, using the perturbation theory, we obtained [5]:

$$b_{mn} = \left(1 - \frac{(Z\alpha)^2}{n^2}\frac{M}{M+m}\right), \ b_{Mn} = \left(1 - \frac{(Z\alpha)^2}{n^2}\frac{m}{M+m}\right), \qquad (7a\text{-}b)$$

$$\gamma_{mn} = \left[1 - \frac{(Z\alpha)^2}{n^2}\frac{M^2}{(m+M)^2}\right]^{-1/2}, \ \gamma_{Mn} = \left[1 - \frac{(Z\alpha)^2}{n^2}\frac{m^2}{(m+M)^2}\right]^{-1/2} \qquad (7c\text{-}d)$$

to the accuracy $c^{-2}$. Here $Z$ is the atomic number, and $\alpha$ is the fine structure constant.

Considering further the radiative corrections to the atomic energy levels, we point out the inconsistency existing in QED of bound states, which is again related to the non-accounting of non-radiating character of electromagnetic fields of electrically bound particles. Namely, the inhomogeneous wave equation for the operator of vector potential

$$\Box \hat{A} = -\frac{4\pi}{c}j, \qquad (8)$$

whose validity is implied in QED of bound states, becomes non-applicable in the stationary energy states. (Here $j$ is the current density, and $\Box$ is the d'Alembert operator). From the historical viewpoint, it is important to stress that the creators of QED could not be aware about this inconsistency, because the presentation of total vector potential $A$ as the sum of bound $A_b$ and radiative $A_r$ components had been achieved only in 70[th] years of the past century (i.e. $A=A_b+A_r$), and neither bound contribution $A_b$ alone, nor radiation contribution $A_f$ alone satisfy eq. (8) [8], i.e.

$$\Box \hat{A}_b \neq -\frac{4\pi}{c}j, \ \Box \hat{A}_f \neq -\frac{4\pi}{c}j. \qquad (9a\text{-}b)$$

At the same time, since the electrically bound particles do not radiate in the stationary energy states, that $\hat{A} = \hat{A}_b$. Hence instead of the equality (8) we get the inequality (9a). However, in further development of QED of bound states this inconsistency acquired a latent form, since eq. (8) is not explicitly used in the diagram technique.

Nevertheless, the indicated inconsistency of QED of bound states anyway must be eliminated, in order to make this theory logically self-consistent, and finally, to be sure that all physical effects are accounted for.

---

[1] We have to notice that particles in the exited energy states have a finite probability to radiate. However, here we assume that the process of radiation (accompanied by the change of energy state of particle) is guided separately by the energy-momentum conservation law, whereas "between" the acts of radiation, the approach of pure bound field remains in force. Anyway, only experiments can validate the correctness of this supposition.

In order to cope with this problem, we assumed that the electromagnetic fields of electrically bound particles in the stationary states are described by the Poisson-like equation [5], which is fulfilled for the bound field component alone. At the same time, the replacement of eq. (8) by the Poisson-like equation does not change anything in the diagram technique, so that in any QED expression, the radiative terms themselves remain unchanged.

Here we remind that many QED expressions for radiative corrections of energy levels of atoms can be presented as the product of two terms: the term issuing from relativistic quantum mechanics and the radiative term. Thus, the pure bound field approach implies that the PBFT corrections emerge only in the former terms. We also notice that in calculation of QED corrections to the atomic energy levels, the approximation of one-body problem (where $M\to\infty$) usually occurs sufficient, and eqs. (6), (7) are simplified as:

$$U \to \gamma_n U, \quad m \to b_n m, \tag{10a-b}$$

with

$$b_n = 1 - (Z\alpha)^2/n^2, \quad \gamma_n = \left(1 - (Z\alpha)^2/n^2\right)^{-1/2} \tag{11a-b}$$

and $\gamma_{Mn}$, $b_{Mn} = 1$.

Thus, the pure bound field corrections to the common results of atomic physics exhibit themselves as a combination of factors (7), or factors (11) at $M\to\infty$. The crucial problem is the comparison of our predictions with the results of measurements, which is done in section 2.

In particular, we subsequently consider 1$S$-2$S$ interval in positronium (sub-section 2.1), 1$S$ hyperfine spin-spin splitting in hydrogen and leptonic atoms (subsections 2.2), the corrections to the Lamb shift in hydrogen and in muonic hydrogen with our re-estimation of the proton charge radius (subsection 2.3). We also analyze the experiments on the measurement of lifetime of bound muon in meso-atoms (subsection 2.4), where the results obtained in the framework of our approach seem important for better understanding of the issue.

It is very important to emphasize that in *all cases* (omitted for brevity in the present paper), where the common results are already in a quantitative agreement with the measurement data (e.g., 1$S$ hyperfine spin-spin interval in muonium, etc.), the pure bound field factors (7) either cancel each other, or give the corrections, lying beyond the measurement precision.

Finally, we conclude in section 3, where we highlight the crucial experiment for further verification of our approach: the new measurements of lifetime of bound muon versus the atomic number $Z$ for various meso-atoms.

## 2. Pure bound field corrections to QED results

In order to demonstrate, how pure bound field corrections are applied in practice, we reproduce from ref. [4] the Breit equation without external field for the Schrödinger-like wave function $\psi(\mathbf{r})$, modified via the replacements (7):

$$\left[\frac{p^2}{2mb_{mn}} + \frac{p^2}{2Mb_{mn}} - \gamma_{mn}\gamma_{Mn}\frac{Ze^2}{r} - \frac{p^4}{8m^3 b_{mn}^3 c^2} - \frac{p^4}{8M^3 b_{Mn}^3 c^2} + U_b(\mathbf{p}_m, \mathbf{p}_M, \mathbf{r})\right]\psi(\mathbf{r}) = W\psi(\mathbf{r}), \tag{12}$$

where $W$ is the energy, and the term $U_b(\mathbf{p}_m, \mathbf{p}_M, \mathbf{r})$ is equal to

$$U_b(\mathbf{p}_m, \mathbf{p}_M, \mathbf{r}) = -\frac{\pi Z e^2 \hbar^2}{2c^2}\left(\frac{1}{b_{mn}^2 m^2} + \frac{1}{b_{Mn}^2 M^2}\right)\delta(\mathbf{r}) - \frac{Ze^2}{2b_{mn}b_{Mn}mMr}\left(\mathbf{p}_m \cdot \mathbf{p}_M + \frac{\mathbf{r}\cdot(\mathbf{r}\cdot\mathbf{p}_m)\mathbf{p}_M}{r^2}\right) -$$

$$\frac{Ze^2\hbar\gamma_{mn}\gamma_{Mn}}{4b_{mn}^2 m^2 c^2 r^3}(\mathbf{r}\times\mathbf{p}_m)\cdot\boldsymbol{\sigma}_m + \frac{Ze^2\hbar\gamma_{mn}\gamma_{Mn}}{4b_{Mn}^2 M^2 c^2 r^3}(\mathbf{r}\times\mathbf{p}_M)\cdot\boldsymbol{\sigma}_M - \frac{Ze^2\hbar\gamma_{mn}\gamma_{Mn}}{2b_{mn}b_{Mn}mMc^2 r^3}\left((\mathbf{r}\times\mathbf{p}_m)\cdot\boldsymbol{\sigma}_M - (\mathbf{r}\times\mathbf{p}_M)\cdot\boldsymbol{\sigma}_m\right) + \tag{13}$$

$$\frac{Ze^2\hbar\gamma_{mn}\gamma_{Mn}}{4b_{mn}b_{Mn}mMc^2}\left[\frac{\boldsymbol{\sigma}_m \cdot \boldsymbol{\sigma}_M}{r^3} - 3\frac{(\boldsymbol{\sigma}_m \cdot \mathbf{r})(\boldsymbol{\sigma}_M \cdot \mathbf{r})}{r^3} - \frac{8\pi}{3}\boldsymbol{\sigma}_m \cdot \boldsymbol{\sigma}_M \delta(\mathbf{r})\right],$$

where $\boldsymbol{\sigma}$ is the Pauli matrix. We point out that with the correcting factors $b_{mn}$, $b_{Mn}$, $\gamma_{mn}$, $\gamma_{Mn}$ equated to unity, eq. (12) transform to the common Breit equation without external field [6].

The details of solution of eq. (12), which can be found in our paper [4], are omitted here for brevity. We notice that this equation gives the same gross and fine structure of atomic energy levels, like the common Breit equation. Further on, we stress the important property of eq. (12): in spite of the fact that the factors (7) differ from unity in the order of magnitude $(Z\alpha)^2$, the pure bound field corrections (expressed as some combination of factors $b_{mn}$, $b_{Mn}$, $\gamma_{mn}$, $\gamma_{Mn}$) to $nS$ levels and fine structure corrections appear only in the order $(Z\alpha)^6 m/M$; the corrections to hyperfine spin-spin interval, as well as the radiative corrections might have the order $(Z\alpha)^2$.

Below we present a short review of application of eq. (12) to the atomic physics, along with QED results modified by the factors (11).

### 2.1. 1S-2S interval in positronium.

As we have mentioned above, the pure bound field correction to $nS$ levels has the order $(Z\alpha)^6 m/M$, and explicitly is equal to [4]

$$\delta W_{PBFT} = \frac{2mc^2(Z\alpha)^6}{n^5(M+m)^5}\left(2mM^4 + m^2 M^3 + 2m^3 M^2\right) - \frac{mc^2(Z\alpha)^6 m^2 M^3}{n^6(m+M)^5} - \frac{15}{8}\frac{mc^2(Z\alpha)^6}{n^6}\frac{mM^2(M^2+m^2)}{(M+m)^5}. \quad (14)$$

One can see that the correction (14) scales as $n^{-6}$ or $n^{-5}$, and for atoms with $m \ll M$ it should be taken into account for 1S state only. For example, for 1S state of hydrogen, the correction (14) is equal to $\delta W_{PBFT}^H(1S) = \frac{9}{8}mc^2(Z\alpha)^6 \frac{m}{M}$ =10.8 kHz, which is comparable with the accuracy of measurement of 1S Lamb shift in hydrogen. Nevertheless, as shown in ref. [4], it gives a minor contribution to the re-estimation of the proton charge radius via the 1S Lamb shift in the hyfrogen atom.

However, the correction (14) becomes significant at $m=M$ (positronium), where for the difference of 1S and 2S states it gives

$$\delta W_{PBFT}^{Ps}(1S-2S) = 2.95 \text{ MHz}, \quad (15)$$

which should be involved into the re-estimation of 1S-2S interval in positronium.

Besides, for positronium the Breit potential includes the additional annihilation term (*e.g.*, [6, 9]), whose pure bound field correction in the orthopositronium case reads [4]:

$$\delta W_{ann} = \frac{mc^2(Z\alpha)^6}{4n^5} = 4.53 \text{ MHz}. \quad (16)$$

Thus the total correction to 1S-2S interval is defined as the sum of eqs. (15), (16):

$$\delta W_{total}^{Ps}(1S-2S) = 7.48 \text{ MHz}. \quad (17)$$

Modern theoretical value of this interval is equal to [10]

$$E_{1S-2S}^{Ps} = 1\ 233\ 607\ 222.2(6) \text{ MHz}, \quad (18)$$

and the most precise experimental result is as follows:

$$(E_{ex})_{1S-2S}^{Ps} = 1\ 233\ 607\ 216(2) \text{ MHz [11]}. \quad (19)$$

One can see that the deviation between the values (18) and (19) more than three times larger than the uncertainty in measurement of 1S-2S interval, and this discrepancy between theory and experiment remained unanswered during a long time.

Now, in accordance with the obtained pure bound field correction (17), we have to decrease the calculated value (18) by 7.48 MHz. Hence the 1S-2S interval in positronium becomes [4]

$$(E_{PBFT})_{1S-2S}^{Ps} = E_{1S-2S} - (\delta W_{total})_{1S-2S}^{Ps} = 1\ 233\ 607\ 214.7(6) \text{ MHz},$$

which already well agrees with the experimental value (19).

In our opinion, this result represents the appreciable achievement of the pure bound field approach.

### 2.2. *1S spin-spin interval in hydrogen and leptonic atoms*

In has been shown in ref. [4] that eq. (12) determines the following expression for spin-spin interval of *nS* levels:

$$(W_{PBFT})_{s-s} = \left(1 - \frac{(Z\alpha)^2}{n^2} \frac{2mM}{(M+m)^2}\right) W_{s-s}, \tag{20}$$

where $W_{s-s}$ is the spin-spin interval calculated in the common approach. At the same time, one should emphasize that eq. (20) does not determine yet the total pure bound field correction, because the term $W_{s-s}$ itself contains the ratios $g_m \sigma_m / m$, $g_M \sigma_M / M$ (where $g_m$, $g_M$ are g-factors for particles *m* and *M*, correspondingly), which, in general, are also the subject of modification in the framework of our approach. Considering below the cases of hydrogen, positronium and muonium, we note that for the firsts two atoms the pure bound field corrections to the mentioned ratios $g_m \sigma_m / m$, $g_M \sigma_M / M$ are negligible in comparison with the term in brackets of the rhs of eq. (20) [12]. Therefore, for the hydrogen atom and positronium, eq. (20) can be directly applied.

In the case of hydrogen eq. (20) gives the numerical value of correction for 1*S* level less than 100 kHz, and can be ignored at the present accuracy of determination of hfs.

For 1*S* state of positronium eq. (20) yields:

$$(W_{PBFT})^{Ps}_{s-s} = \left(1 - \frac{(Z\alpha)^2}{2n^2}\right) W^{Ps}_{s-s}, \tag{21}$$

where
$$W^{Ps}_{s-s} = 203\ 391.7(8) \text{ MHz} \tag{22}$$

is the value presently calculated [10].

The corresponding experimental data are 203 389(2) [13] and 203 387(2) [14], which disagree with the value (22) by ~2 standard deviations.

We stress that the mentioned deviation between calculated and experimental data with respect to 1*S* spin-spin interval in positronium remained puzzling during a long time.

Now eq. (21) allows us to compute the corrected hyperfine spin-spin interval in positronium, using the numerical value (22):

$$(W_{PBFT})^{Ps}_{s-s} = 203\ 386(1) \text{ MHz}. \tag{23}$$

This result is already in a good agreement with the experimental data and, in our opinion, it represents one more appreciable achievement of pure bound field approach.

Finally, applying eq. (20) to the case of muonium, we have to involve additionally the pure bound field correction to the magnetic moment-to-mass ratio for muon. This ratio is determined experimentally via the Zeeman effect. For the case of muonium we obtain the following expression of for the Zeeman splitting of energy levels [12]:

$$(W_{PBFT})^{Mu}_{Magnetic} = \left(1 - \frac{(Z\alpha)^2}{n^2} \frac{2mM}{(M+m)^2}\right) W^{Mu}_{Magnetic}, \tag{24}$$

where $W^{Mu}_{Magnetic}$ is the value calculated in the framework of the common approach, and $M = m_\mu$ now stands for the muon mass. Since $W^{Mu}_{Magnetic}$ is linearly proportional to the magnetic moment, we get the related re-estimation of the magnetic moment of muon $\mu_\mu$:

$$(\mu_\mu)_{PBFT} = \left(1 - \frac{(Z\alpha)^2}{n^2} \frac{2mM}{(M+m)^2}\right) (\mu_\mu)_c, \tag{25}$$

where $(\mu_\mu)_c$ is the magnetic moment of muon calculated in the common theory.

Further we observe that the energy $W_{s-s}$ is linearly proportional to magnetic moment, and introducing the pure bound field correction (25), we find that the correcting factors in eq. (20) and eq. (25) (i.e. the terms in the bracket of rhs) exactly cancel each other, so that the total correction to hfs interval in muonium disappears.

It is known that the value of 1$S$ spin-spin interval in muonium calculated in the common approach perfectly agrees with the measurement data (see, e.g. [10]). This result represents the particular manifestation of our statement made in the introduction section: in all cases, where QED results are already in a quantitative agreement with the measurement data, the correcting factors (7) either cancel each other, or give the values of corrections, lying beyond the measurement precision.

We can add that for muonic hydrogen, the nuclear size effect is much larger than any pure bound field corrections, resulting from eq. (20) and re-estimation of magnetic moment of muon. In particular, one can show that the re-estimation of Zemach proton radius via hfs in muonic hydrogen leads to the difference from the common value about $1.0 \cdot 10^{-3}$ fm [12], which is smaller than the present uncertainty in determination of this parameter.

2.3. *Corrections to the Lamb shift.*

In the analysis of radiative corrections to the atomic energy levels, where the approximation of one-body problem is well-fulfilled, the pure bound field modifications (10) in the input of QED expressions should be accounted for. Below we present the corrections to the Lamb shift $L$ for light hydrogenlike atoms, which emerge in the framework of our approach. In particular, for the 2$S$-2$P$ Lamb shift we obtain [4]

$$(L_{PBFT})_{2S-2P} = \gamma^2_{n=2} L_{2S-2P} = L_{2S-2P} \left(1 - (Z\alpha)^2/4\right)^{-1}, \qquad (26a)$$

where $L_{2S-2P}$ is the value of Lamb shift calculated in QED, and we have used eq. (11b) at $n$=2.

Thus the relative pure bound field correction to the 2$S$-2$P$ Lamb shift is equal to

$$\frac{\delta L_{2S-2P}}{L_{2S-2P}} = \frac{(L_{PBFT})_{2S-2P} - L_{2S-2P}}{L_{2S-2P}} = \left[\left(1 - (Z\alpha)^2/4\right)^{-1} - 1\right] = 1.33 \cdot 10^{-5} \text{ (at } Z = 1\text{)} \qquad (26b)$$

The correction (26b) substantially exceeds the relative measurement precision of the 2$S$-2$P$ Lamb shift for Doppler-free two-photon laser spectroscopy [15] and in the case of hydrogen, it is comparable with the relative nuclear structure contribution. Therefore, the correction (26b) does affect the value of the proton charge radius $r_E$ calculated via the classic Lamb shift.

In order to determine the proton size resulting from our approach, it is sufficient to use the quadratic parametrization for the 2$S$-2$P$ Lamb shift, i.e.

$$L_{2S-2P} = A + B r_E^2, \qquad (27)$$

where $L_{2S-2P}$ stands for the measured value, and in the hydrogen case the coefficients $A$ and $B$ are equal to [16]

$$A=1057695.05 \text{ kHz}, \quad B=195.750 \text{ kHz/fm}. \qquad (28\text{a-b})$$

Due to eq. (26a), in our approach eq. (27) is modified to the form

$$L_{2S-2P} = \gamma^2 A + \gamma^2 B (r_{PBFT})_E^2, \qquad (29)$$

and combining eqs. (27) and (29), we derive the relationship between the proton charge radius, evaluated in our approach $(r_{PBFT})_E$, and the commonly adopted value $r_E$:

$$(r_{PBFT})_E = \sqrt{\frac{r_E^2}{\gamma^2} + \frac{A(1-\gamma^2)}{B\gamma^2}} = \sqrt{\frac{r_E^2}{\gamma^2} - (Z\alpha)^2 \frac{A}{4B}}, \qquad (30)$$

where we have taken into account eq. (11b) for factor $\gamma$ at $n$=2.

This equation shows that in pure bound field theory the proton charge radius should be *smaller* in comparison with its common evaluation. Substituting into eq. (30) the numerical values (28) along with the CODATA-2010 value $r_p$=0.8775(51) fm [3], we obtain.

$$(r_{PBFT})_E = 0.834(6) \text{ fm.}$$

This estimation is already much closer to the proton size derived in refs. [1, 2], than the CODATA value of $r_E$. At the same time, we recall that the CODATA value of the proton charge radius incorporates the experimental data in both particle physics and atomic physics, and, in general, is less than the proton size derived from the classic Lamb shift solely. In particular, the modern data on 2S-2P Lamb shift in hydrogen obtained by various authors within the common approach (see refs. [10, 16] and references therein) define the range of variation of the value of $r_E$ between 0.875 fm and 0.891 fm. Thus taking the midpoint $r_E$ =0.883 fm, we obtain

$$(r_{PBFT})_E = 0.841(6) \text{ fm}, \qquad (31)$$

which exactly coincides with the new measurement of the proton size [1, 2].

We add that for muonic hydrogen, the relative pure bound field correction to 2S-2P Lamb shift (26b) occurs much less than the relative contribution of the nuclear size effect. Thus, this correction practically does not influence the proton size evaluated in muonic hydrogen via the common approach and via pure bound field approach. In particular, using the parameterization (1) of ref. [2] with the numerical coefficients multiplied by $\gamma_{n=2}^2$ times according to eq. (26a), we derive the difference between $(r_{PBFT})_E$ and $r_E$ about $3 \cdot 10^{-4}$ fm, which is less than the uncertainty of the muonic experiment [2].

Thus, the exact coincidence of the proton size obtained via the classic Lamb shift and laser spectroscopy of muonic hydrogen represents, in our opinion, a very important achievement of pure bound field approach.

What is more, for the 1S Lamb shift in hydrogen, our approach also predicts the decreased proton size in comparison with the CODATA value. Omitting particular calculations, which can be found in ref. [4], we present the final result of re-estimation of $r_E$ via the 1S Lamb shift: $(r_{PBFT})_E$ =0.846(22) fm. Though this value has a much larger uncertainty that the results of calculation of the proton size in the classic Lamb shift experiments and in muonic hydrogen spectroscopy, it well agrees with both of them.

### 2.5. *Lifetime of bound muon in meso-atoms*

In this sub-section we separately consider one more effect predicted in the framework of our approach and which, as we will show below, opens a new possibility for the experimental verification of our predictions.

Namely, we pay attention on the fact that the replacement (10b) $m \to b_n m$ for the quantum one-body problem simultaneously implies the replacement $E = \gamma_n mc^2 \to \gamma_n b_n mc^2$ in the expression for the energy of bound particle, related to its motion. Due to the direct relativistic relationship between the quantities "energy" and "frequency" (or "time rate"), the mentioned modification of energy signifies that the time rate $t'$ for bound particle is also modified by $\gamma_n b_n$ times in comparison with the laboratory time $t$, i.e.

$$dt' = dt/b_n \gamma_n . \qquad (32)$$

The additional argumentation in the favor of eq. (32) can be obtained in the analysis of ways of solution of the Dirac-Coulomb equation (for one-body problem) and Breit equation (for two-body problem) modified in our approach; for more details see ref. [17].

The physical meaning of eq. (32) can be clarified in the semi-classical limit of the one-body problem, where the coefficient $b$ is defined by eq. (3c) at $M=\infty$. Thus, eq. (32) shows that the electric binding energy $U$ affects the time rate for the orbiting bound electron, though we stress that this effect is quantum in its origins, and is not extended to the classical case [17].

The convenient objects for verification of eq. (32) are meso-atoms with different atomic number $Z$, where the negative muons being captured by the atoms and reaching 1S state possess a property to directly exhibit their time rate via the lifetime $\tau_b$. Since the electric binding energy $U$ is linearly proportional to the atomic number $Z$, we get a unique possibility to verify eq. (32) via the measurement of lifetime $\tau_b$ of bound muons as the function of $Z$ in various meso-atoms.

The experiments for measurement of lifetime of muons bound in meso-atoms at various $Z$ had been carried out in 1960's of the last century [18, 19] and their results at large $Z$ contradict to each other, as well as to the most reliable theoretical predictions made by Huff [20], see Fig. 1.

Chronologically, the experiment by Yovanovitch [18] was implemented before the experiment by Blair et al. [19]; moreover a drastic deviation of experimental data of [18] (black points) at large $Z$ from the careful calculations by Huff [20], stimulated the authors of [19] to carry out new measurements on this subject. The results obtained in [19] are shown in Fig. 2 as the hollow circles. One can see that at large $Z$, these results are in a good agreement with the idealized curve by Huff (thin continuous line in Fig. 1).

Thus, after the implementation of the experiment [19], it was commonly decided that the data by Yovanovitch [18] (black points in Fig. 1) are most likely erroneous, and the entire problem had been supposed to be closed.

However, it was fully forgotten that the experimental data *must be compared* not with the idealized curve by Huff (thin continuous line in Fig. 1), but rather with his real curve (bold continuous line in Fig. 1), which is obtained through the corrections of the idealized curve to the difference of electron spectra for bound and free muons, as well as to a finite size of target [20]. We see that with respect to the real curve, both the Yovanovitch data [18], and Blair et al. data [19] give deviating results. Thus, a crucial question: whose experimental data are incorrect – by Yovanovitch or Blair et al. – remains unanswered.

In the framework of our approach we have to adopt that the real curve by Huff (bold continuous line in Fig. 1) is still incorrect, because it does not take into account the change of time rate (32) for a bound muon. More specifically, the calculations of Huff include the relativistic dilation of time for a muon [20], expressed in eq. (32) by factor $1/\gamma_n$. Hence, our correction to the calculated value of the time rate of bound muon is presented by factor $1/b_n$. Since the observed lifetime of muon is inversely proportional to its time rate, we get the relationship

$$(\tau_{PBFT})_b = b_n (\tau_{Huff})_b, \qquad (33)$$

where $(\tau_{Huff})_b$, as the function of $Z$, is presented in Fig. 2 as the bold continuous line.

Thus, using eq. (33) we multiply the Huff data by factor $b_n$, defined by eq. (11a) at $n=1$. As the outcome, we obtain the corrected dependence $(\tau_{PBFT})_b$ on $Z$ to be shown in Fig. 1 as the dot line [17]. A similar curve has been obtained in the earlier paper by the third author [21].

We see that at large $Z$ the PBFT curve is in a good agreerment with the data by Yovanovitch, and this coincidence makes highly unbelievable that the assumed effect (32) and the measurements by Yovanovitch are both wrong.

Therefore, for the clarification of the enite issue, new experiments for measurement of lifetime of bound muon are highly required, especially for meso-atoms with a large $Z$.

## 3. Conclusion

In a light of the present discussion on the proton size, stimulated by the recent measurements [1, 2], we address to the approach of pure bound field theory, which we recently suggested [4, 5] for electrically bound quantum particles.

In the present contribution we presented a brief review of application of pure bound field approach to precise physics of simple atoms, and demonstrated the undoubted successes of this approach in the elimination of long-standing discrepancies between theory and experiment.

With respect to the present discussion about the proton size, the re-calculated values of the proton charge radius $r_E$ are as follows:

2$S$-2$P$ classic Lamb shift in hydrogen  $r_E$=0.841(6) fm;
1$S$ Lamb shift in hydrogen  $r_E$=0.844(22) fm;
laser spectroscopy of muonic hydrogen  $r_E$=0.84087(39) fm.

We find the fact of coincidence of the values of $r_E$, obtained in three independent kinds of measurments, to be very important. Thus we believe that further experimental verification of pure bound field approach represents a topical problem, and in the present paper we suggested an experiment: the new measurement of lifetime of bound muon in various meso-atoms, especially at large *Z*, where the standard calculations and our predictions essentially deviate from each other.

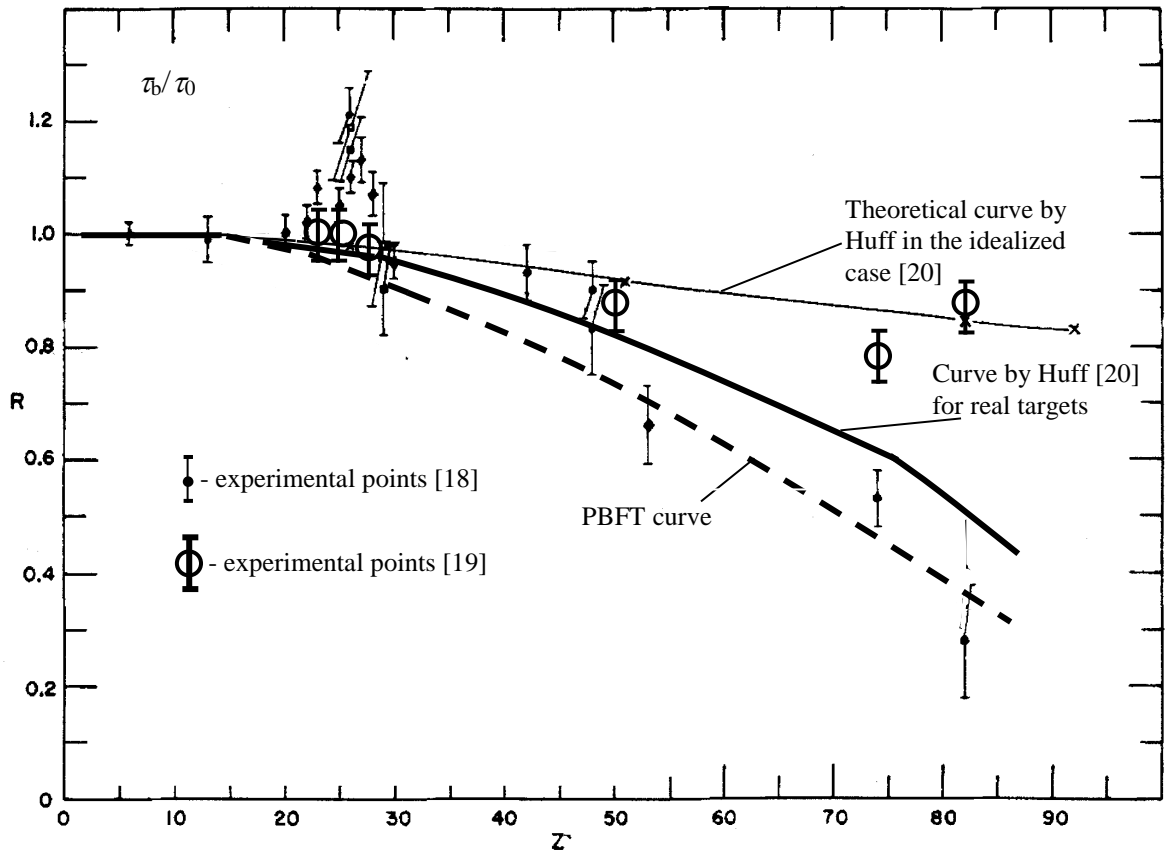

Fig. 1. The dependence of bound muon lifetime $\tau_b$ on $Z$. We compare the results of theoretical calculations by Huff [20] (continuous lines) corrected in pure bound field theory PBFT (dot line) with the experimental data of [18] and [19].